\documentclass[useAMS,psfig]{mn2e}
\input{psfig}
\def\j{2QZ215454.3$-$305654 }

\title[A radio-quiet BL Lac object]{2QZJ215454.3-305654: a radio-quiet BL Lac object or lineless QSO?}
\author[D.Londish et al.]
	{D. Londish$^{1,2}$, J. Heidt$^{3}$, B.J. Boyle$^{4,2}$, S.M. Croom$^{2}$, L. Kedziora-Chudczer$^{4,1}$\\
${^1}$ University of Sydney, School of Physics, Sydney NSW 2006, Australia \\
${^2}$ Anglo-Australian Observatory, PO Box 296, Epping, NSW 1710, Australia\\
${^3}$ Landessternwarte Heidelberg, K\"onigstuhl, D$-$69117 Heidelberg, Germany\\
${^4}$ Australia Telescope National Facitity,PO Box 76, Epping NSW 1710, Australia\\}

\begin{document}

\pagerange{\pageref{firstpage}--\pageref{lastpage}} \pubyear{2004}

\maketitle

\label{firstpage}

\begin{abstract}
High signal-to-noise spectroscopy  has established a redshift of $z=0.494$ for the source 2QZJ215454.3$-$305654, originally selected from the 2dF/6dF QSO Redshift Surveys as one of 45 candidate BL Lac objects displaying a  featureless continuum at optical wavelengths. Radio observations using the  Australia Telescope Compact Array at 1.4 GHz place a 3$\sigma$ upper limit on the object's radio flux density of $\sim$0.14mJy. The   radio-to-optical flux ratio of this object is thus more than 7 times
lower than  the lowest such ratio observed in BL Lac objects.  While the optical properties of 2QZJ215454.3$-$305654 are consistent with a BL Lac identification, the lack of radio and/or X-ray emission is not. It is uncertain whether this object is an AGN dominated by optical continuum emission from an accretion disk, or is similar to a BL Lac object with optical nonthermal emission from a relativistic jet.

\end{abstract}

\section{Background}

The candidate BL Lac object 2QZJ215454.3$-$305654 was observed in August 2003 using the VLT\footnote{VLT-UT4 on Cerro Paranal (Chile) operated by the European
  Southern Observatory in the course of the observing proposal
  71.A-0174.}. It was previously observed as a colour-selected point source 
in June 2000 as part of the 2dF QSO Redshift Survey (2QZ, Croom et al. 2001; 2004). By examining spectra of point sources from both the 2dF and 6dF Redshift Surveys (2QZ/6QZ) Londish et al.\,(2002) began a campaign to identify the first optically selected  BL Lac sample, the 2BL, unbiased with respect to the objects' radio and X-ray flux densities. 2QZJ215454.3$-$305654 was identified as one of 45 such featureless continuum objects.  

Subsequent cross-correlation of these 45 candidate BL Lacs with the NVSS/FIRST radio catalogues\footnote{NRAO/VLA Sky Survey (Condon et al.\,1998) and Faint Images of the Radio Sky at 20cm (White et al.\,1997)} produced  matches for nine sources only; five of these objects are also detected in the RASS X-ray catalogue\footnote{{\it ROSAT} Bright and Faint All Sky Survey X-ray catalogues (RASS, Voges et al.\,1999)}. No detection was found in either NVSS or RASS  at the coordinates of 2QZJ215454.3$-$305654. The optical apparent magnitudes of the 2BL cover the  range  $18.25 < b_{\rm J} < 19.97$, equating to $\sim$0.2--0.04mJy;  radio flux densities $>$ 1mJy would thus be expected for at least 50\% of the sample if these objects are similar to hitherto detected BL Lacs (i.e.\,radio loud with a ratio of radio-to-optical flux $>$ 10). 

The average signal-to-noise ratio (SNR) of the 2dF/6dF spectra in the 2BL is low (10--15).  To eliminate contamination of the sample by objects such as featureless Galactic White Dwarfs we carried out proper motion studies based on data from the updated (June 2003) SuperCosmos Sky Survey\footnote {SuperCosmos Sky Survey at www-wfau.roe.ac.uk/sss, maintained by the Institute for Astronomy,
Royal Observatory, Edinburgh}. We also investigated the spectral energy distribution (SED) of 2BL objects from the 2dF equatorial strip\footnote{The 2QZ survey covers $\sim$740 deg$^2$ of sky, comprising  two 
75$^{\rm o} \times  5^{\rm o}$ strips, one
centered on $\delta =  0^{\rm o}$ with RA
range 9$^h$50 to 14$^h$50 (equatorial strip) and the other on $\delta =  -30^{\circ}$ with RA range 21$^h$40 to 3$^h$15
(southern strip)}  at optical and infrared wavelengths using data from the Sloan Digital Sky Survey (SDSS, Early Data Release, Stoughton et al.\,2002) as well as our own $J$- $H$- and $K$-band observations using the 2.1m telescope at Kitt Peak (Londish 2003, Londish et al.\,2004 in preparation).  We also examined the 2dF/6dF spectra for evidence of intervening absorbers (Outram et al.\,2001)
 as a further means of distinguishing between a Galactic WD and an extragalactic object; no intervening systems were found for 2QZJ215454.3$-$305654.

In this paper we report in \S\,2 on follow-up observations of 2QZJ215454.3$-$305654 and discuss in \S\,3 the implications of our findings with respect to the identity of this object. Throughout the paper we use $\Omega_{\rm m}=0.3$, $\Omega_{\rm \Lambda}=0.7$, $H_0
=70\,$km$\,$s$^{-1}$Mpc$^{-1}$, and define spectral index as $f_{\nu} \propto \nu^{-\alpha}$.\

\section{Observations of 2QZJ215454.3$-$305654}

\subsection{Proper motion studies}

Originally 78 objects were selected from the 2QZ/6QZ surveys as being featureless continuum objects. Inspecting the SEDs of 24 objects found in SDSS and  our own infrared observations at Kitt Peak we determined that a proper motion cut at 2.5$\sigma$ (where 1$\sigma \simeq 15$ mas yr$^{-1}$) provided an effective discrimination between objects likely to be Galactic white dwarfs (characterised by  high proper motion and thermal SEDs in the optical/IR bands) and potential extragalactic objects (non-thermal SEDs and no detectable proper motion). Removing objects with high proper motion resulted in the final sample of 45 candidate BL Lac objects, including 2QZJ215454.3$-$305654
with a measured proper motion of 7.5 $\pm 16.3$ mas yr$^{-1}$. The $u' - b_{\rm J}$ and $b_{\rm J} - r$ colours of this object in the 2QZ system (Smith et al.\,2004) are $-$0.78 and 0.66 respectively.
\begin{figure}
\psfig{file=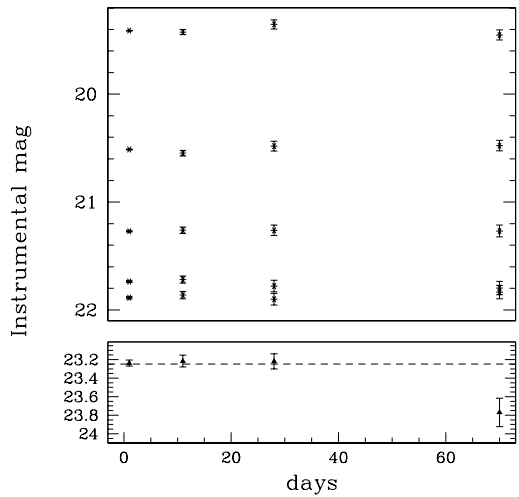,height=2.8in}
\caption{ Top panel: Instrumental magnitudes (arbitrary zero point) of 5 stars; Lower panel:  Instrumental magnitudes for 2QZJ215454.3-305654 using the 74'' telescope Aug 2001--Oct 2001.  A reduced $\chi^2$ of 4.0 from a line of best fit (dashed line) was calculated for the 2BL object, indicative of a varying source.}
\label{2154}
\end{figure}
\subsection{Variability Studies}

2QZJ215454.3$-$305654 was observed in $B$-band using the MSSSO 74'' telescope at Mt Stromlo, ACT,  Australia, between August and October 2002 as part of a year long campaign to look for evidence of variability in the 45 2BL objects. In analysing the data instrumental magnitudes of the candidate BL Lac object were compared to those of five other stars in the same field. Conditions were not photometric and no attempt was made to convert from instrumental magnitudes to apparent magnitudes. Instead magnitude differences between the  first and subsequent observations of the non-varying stars were calculated, and an average shift between images computed. This correction was subsequently applied to all objects in the image so that instrumental magnitudes
were normalised to that of the first observation; relative changes in magnitude of the candidate BL Lac object with respect to the field stars could thus be observed. Any stellar object found to be varying by more than 0.05 mag was rejected and the shifts recalculated.\
2QZJ215454.3$-$305654 was found to vary by up to 0.55 mag (Fig.\,\ref{2154}).

\begin{figure}
\psfig{file=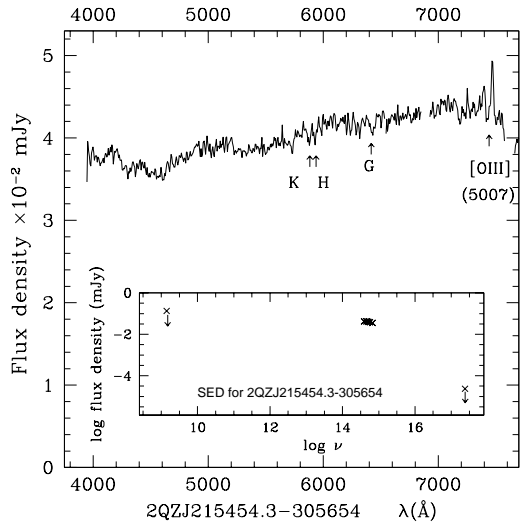,height=3.0in}
\caption{Spectrum of 2QZJ215454.3$-$305654 taken at the VLT showing Ca\,II H \& K and G-band absorption features and [OIII]\,$\lambda$\,5007 emission at $z = 0.494$. Telluric absorption features at 6540, 6860--6920 and 7585--7655 \AA\ have been removed. Insert shows SED with upper limits for X-ray and radio flux (x-axis is observed frequency).}
\label{vlt}
\end{figure}

\subsection{Spectroscopy}

A low-resolution spectrum of \j was recorded on the night of August 26/27 2003 with FORS2 at the ESO 8m VLT-UT4, Paranal, Chile. We used grism 150I, which gave us a spectral scale of $\sim$6.9 \AA\, per pixel. The slit width was set to 1'' and
integration time was 900 sec. The effective wavelength range covered was 
$\sim$4000$-$7800\,\AA\ (longward of 7800\,\AA\ the spectrum is heavily contaminated by night sky emission and second order overlap). During the night the 
spectrophotometric standard  LTT 7379 from Oke (1990) was observed.

The data reduction of the spectrum (bias subtraction, flatfielding, cosmic ray removal, sky subtraction, wavelength and flux calibration, etc.) was performed using standard IRAF routines. The FWHM spectral resolution measured from strong night sky emission lines is $\sim 25\ {\rm \AA}$. The spectrum is shown in Fig\,\ref{vlt}; several features ---  Ca\,II H \& K and G-band absorption, and a weak emission feature consistent with [OIII]\,$\lambda$\,5007 ---  were detected at a redshift of 0.494 (see Table \ref{lines}) It is also possible to see a weak emission feature at 7253\AA\ consistent with H$\beta$($\lambda$4861), while the features at 5730\,\AA\ and 5931\,\AA\ might be indicative of the Balmer lines H$\eta$ (3835\AA) and  H$\epsilon$ (3970\AA) in absorption. However we note that the lines of Hgamma and Hdelta are not seen
in the spectrum, making these higher order Balmer series line
identifications less certain. \
\begin{table}
\begin{center}
\begin{tabular}{|l|c|c|c|} \hline
\multicolumn{1}{|c|}{} &
\multicolumn{1}{|c|}{rest $\lambda$ (\AA)} &
\multicolumn{1}{|c|}{measured $\lambda$} &
\multicolumn{1}{|c|}{$z$}\\
\multicolumn{1}{|c|}{} &
\multicolumn{1}{|c|}{} &
\multicolumn{1}{|c|}{($\pm 4$ \AA) }&
\multicolumn{1}{|c|}{} \\
\hline
Ca\,II K & 3933 & 5877 & 0.4943\\
Ca\,II H & 3968 & 5929 & 0.4942\\
G band & 4303 & 6430 & 0.4943 \\
${\rm [OIII]}$ & 5007 & 7475 & 0.4929\\
H$\beta$ & 4861 & 7254 & 0.4923\\
\hline
\end{tabular}
\end{center}
\caption{Measured wavelengths and deduced redshifts}
\label{lines}
\end{table} 

\begin{figure}
\psfig{file=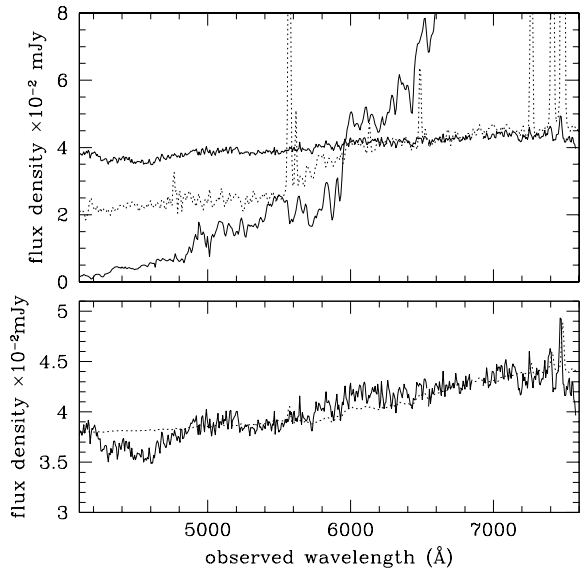,height=3.3in}
\caption{Upper panel: spectrum of 2QZJ215454.3$-$305654 (solid line) compared to an elliptical (solid line) and starburst template (dotted line), scaled to the same flux as 2QZJ215454.3$-$305654 at 5960\AA. $\,$ Lower panel: spectrum of 2QZJ215454.3$-$305654 (solid line) compared to a composite starburst + elliptical + flat spectrum power-law (dotted line) in the ratio 0.015:0.045:0.94 }
\label{test}
\end{figure}

From the slope in Fig.\,\ref{vlt} the optical spectral index (4500--7500 \AA) is calculated to be $f_{\nu} \propto \nu^{-0.36}$, typical of blue Type I AGN (e.g. QSOs). The CaII `break 
contrast', defined as the relative depression of the 
continuum blueward of the  Ca{\sc ii} H and K absorption lines (namely
 $Br_{4000} = \frac{f^{+} - f^{-}}{f^{+}}$, where $f^{+}$ is 
the average flux at 4050--4250 \AA, and $f^{-}$ is average flux at 
3750--3950 \AA), is $\sim 0.02$, indicative of a strong nonstellar continuum. Inspection of galaxy templates  in the online Kinney-Calzetti Spectral Atlas of Galaxies\footnote{ftp://ftp.stsci.edu/cdbs/cdbs2/grid/} (Calzetti et al.\,1994) shows the spectrum of \j to be inconsistent with that of either a  starburst or bulge/elliptical (Fig.\,\ref{test}, upper panel). However, a composite using a contribution of 1.5\% starburst and 4.5\% elliptical gives a good approximation of emission line strength and the rise in flux at the 4000\AA\ break seen in the spectrum of  2QZJ215454.3$-$305654.  Using starburst and elliptical spectral templates plus  a flat spectrum power-law  in the ratio 0.015:0.045:0.94 (at 5960 \AA) also
reproduces the blue continuum longward of 3500\AA\ restframe (Fig.\,\ref{test}, lower panel).  This provides evidence for a non-stellar origin to the blue component, i.e. either thermal emission from an accretion disk or synchrotron emission from a jet.  \

Unfortunately  no infrared photometry is available for 2BL objects in the southern strip of the 2QZ. Non detection of the source by RASS places  an upper limit on the X-ray emission of $8 \times
10^{-14}$\,erg\,s$^{-1}$\,cm$^{-2}$ (2.36 $\times 10^{-5}$mJy). The overall spectral energy distribution (SED) of \j is shown in the insert to Fig.\,\ref{vlt} using upper limits from RASS and our own observations at 1.4GHz for the radio.\ 

\begin{table*}
\caption[]{Results of the fits to \j}
\begin{tabular}{l|ccccccccc}
\hline
& & & & & & & & & \\
Fit & $\chi^2$ & $m_{core}$ & $M_{core}$ & $m_{gal}$ & $M_{gal}$ & $r_e$    
& $r_e$ & $\epsilon$ & PA\\
    &         &  [mag]   &  [mag] & [mag] & [mag]  
& [arcsec] & [kpc]&            & [deg]\\
& & & & & & & & &\\
\hline
& & & & & & & & &\\
AGN      & 7.21 & 18.83$\pm$0.03 & -22.58 &       &        &      &     &      & \\
Ell.     & 2.39 &                &        & 18.59$\pm$0.06 & -23.27 & 0.13$\pm$0.04 & 0.5 & 0.22 & 175\\
Disk     & 3.68 &                &        & 18.71$\pm$0.05 & -22.95 & 0.18$\pm$0.03 & 0.6 & 0.27 & 171\\
Ell.+AGN & 1.25 & 19.29$\pm0.05$ & -22.12 & 19.08$\pm$0.04 & -22.78 & 0.73$\pm$0.13 & 2.5 & 0.27 & 179\\
Disk+AGN & 1.22 & 19.07$\pm$0.04 & -22.34 & 19.56$\pm$0.04 & -22.10 & 0.83$\pm$0.11 & 2.9 & 0.24 & 179\\
\end{tabular}
\end{table*}

\subsection{Morphological analysis}

\begin{figure}
{\hspace*{1.0cm}{\psfig{file=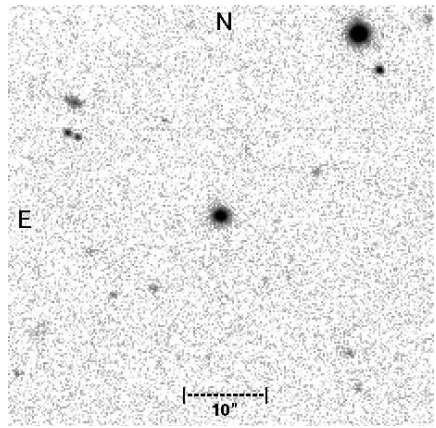,height=2.2in}}}
\caption{VLT acquistion image of 2QZJ215454.3$-$305654 (centre); grey scale is logarithmic and shows a slight E-W elongation of the source, compared to a star to the NW.}
\label{acim}
\end{figure}

To study the morphology of our target, we used two VLT $I$-band 
images  of \j lasting 15 sec each. These were taken as aquisition for 
the spectroscopy, and not specifically for host galaxy studies. After standard data reduction, the two images were aligned
and summed (Fig.\,\ref{acim}). Photometric calibration was performed using Landolt 
standard stars (Landolt 1992).

Although the `break contrast' observed at $\sim 6000$\,\AA\ is virtually zero,  and  results  of template fitting described in \S2.3 suggest a dominant core, the combined 30\,sec $I$-band image   of \j shows evidence of a host galaxy in the slightly elongated shape of the target source. A similar absence of a detectable break contrast is observed in the BL Lac 1ES\,1853+671 (Perlman et al.\,1996) even though the host galaxy is resolved (Heidt et al.\,1999). We therefore proceeded with a fully 2-dimensional
decomposition of this source using the routines  
described in Nilsson et al.\,(1999). 
We fitted five different models to \j -- 
one representing an AGN (scaled PSF), two representing
a galaxy (elliptical or disk-type galaxy) and two representing an AGN + galaxy
(scaled PSF + elliptical/disk-type galaxy). 
The core is described by three parameters ($x_c$, $y_c$, $m_c$), while the 
galaxy is described by seven parameters ($x_g$, $y_g$, $m_g$, $r_e$, $\epsilon$,
PA and shape parameter $\beta$ = 0.25 for an elliptical and $\beta$ = 1 for a
disk type galaxy). For the AGN+galaxy fits we did not allow for an offset
between the centres of the AGN and galaxy. The galaxy models were convolved
with the PSF. 
To extract the PSF, several non-saturated stars in the field
brighter than \j itself were combined.

The results of our fits are summarized in Table 2. 
$K$-corrections were derived from Fukugita et al.\,(1995). 
We used
$K$(I) = 0.45 mag for the elliptical galaxy and 0.25 mag for the disk-type
galaxy (Sbc). No $K$-corrections were applied to the AGN (we assumed
a power-law spectrum of the form $I_{\nu} \propto \nu^{-1}$). 
Extinction (0.05 mag) was taken from NED\footnote{NASA/IPAC Extragalactic Database, which is operated by the Jet Propulsion Laboratory, California Institute of Technology, under contract with the National Aeronautics and Space Administration} To make the half-light
radius derived  for \j comparable to the ones derived for the 
$\epsilon$ = 0 fits in other studies, we multiplied the half-light radius in
arcsec by $\sqrt{(1-\epsilon)}$ when converting to kpc. 

As can be seen from Table 2, neither a pure core-fit nor a pure galaxy-fit
gave good results. The best fit was obtained by a combination of a
core point source + a galaxy. We note that the host galaxy and core are of similar brightness according to our 2-D fitting, contrary
to what might be expected from the spectrum of \j and the simulated composite spectrum in Fig.\,3 which in $I$-band has a power-law contribution of 85\%. This can be explained by the factthat slit losses for the host galaxy are more serious than 
for the point source since the host
galaxy profile is much more extended and flatter with respect to the
central point source. Simulations reveal that slit loss for the core is about 0.25 mag, whereas for a bulge galaxy it would be 0.9 mag. This factor of two ($\sim 0.75\,$mag) is nevertheless insufficient to fully explain the observed difference between the imaging and 
spectroscopic results regarding the relative contribution between the host and core.  Any remaining discrepency, however, is consistent with the errors in the fitting procedures (both imaging and spectroscopic).   

In the image fitting no one host galaxy type is preferred 
over the other, however the  integration time for the aquisition image was very short  (2 $\times$ 15s). Almost all  
known BL Lacs have been found to reside in an elliptical host, with the notable exception of PKS\,1413+135, which appears to reside in an edge-on spiral (M$^c$Hardy et al.\,1994). 
The luminosity of the active nucleus and of the host galaxy  are both in the
lower range found for BL Lac objects at z $\sim$ 0.5 (see Heidt et al.\,2004).

\subsection{Polarisation Studies}

Polarisation studies of a subsample of the featureless continuum objects selected from the 2QZ were conducted in April 2003 using the VLT FORS1 in service observing mode (see Kedziora-Chudczer et al., in preparation). Data reduction was
carried out using software developed by
Jeremy Bailey based on the Starlink POLPACK package.

All BL Lacs discovered to date have detectable optical
polarisation, typically 2-15\% for low-luminosiy,
X-ray selected BL Lacs (Jannuzi et al. 1994). For
2QZJ215454.3$-$305654 only 0.5\% $\pm 0.2$\% linear polarisation was measured in
R-band. These measurements may have been made during the minimum
polarization state of this source, given that polarisation of BL Lacs is
known to be variable and can at times be too low to be detectable (Fan et al.\,2001;  Fan \& Lin 2000).

\subsection{Radio observations at 8.4 and 1.4 GHz}
 
Deep radio observations  at 8.4 GHz were conducted of a subsample of 2BL objects using the Very Large Array (VLA) radio telescope, Socorro, NM. 
2QZJ215454.3$-$305654 was observed in January 2001 using the VLA in a hybrid BnA$\rightarrow$B configuration with 25
antennas operating. Two 20-minute observations were taken using a single frequency channel.  Flux densities were bootstrapped from the calibrator 3C286, however the maximum elevation of the secondary calibrator (2151-304) was  24$^o$, compared to 65$^o$  for the primary. \
\begin{figure}
\psfig{file=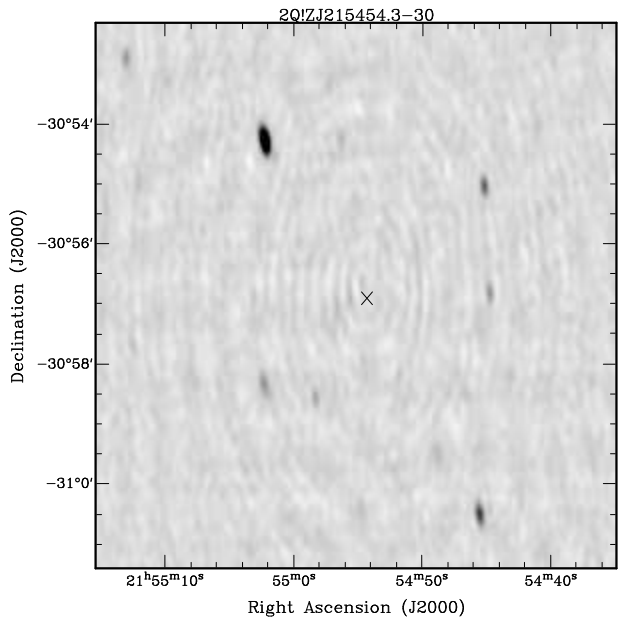,height=3.2in}
\caption{1.4 GHz ATCA radio map; the ATCA field centre was offset 15arcsec from
the optical position of of 2QZJ215454.3$-$305654 is marked with an $\times$.}
\label{atca}
\end{figure}

Data were reduced using the ATNF's\footnote{Australia Telescope National Facility} \textsc{miriad} software 
package (Sault, Teuben \& Wright 1995), following standard procedues from the Miriad User's Manual (Sault and Killeen 1999). The rms noise in the cleaned image is
$\sim 100 \mu$Jy; there is no detectable 8.4 GHz flux  at the coordinates of our candidate BL Lac. 

Observations of 2QZJ215454.3$-$305654 at 1.432 and 1.344 GHz ($2 \times 128$ MHz bandwidth) were carried out in November 2003 using the Australia Telescope Compact Array (ATCA). The array was in the 1.5D configuration with baselines ranging from 107m to 4439m, however the six shortest baselines suffered from solar contamination thus only the 9 longest baselines ($>$ 1km) were used in creating the radio map. The source was observed for 22 $\times$ 40 min over two days. Flux densities were bootstrapped from the primary calibrator PKSB1934-683. Data reduction was again performed using \textsc{miriad}. There is no detectable radio emission at the coordinates of our source above the rms noise of $\sim 45 \mu$Jy (see Fig.\,\ref{atca}). This places a 3$\sigma$ upper limit of 135 $\mu$Jy on the 1.4GHz radio emission of this source, equating to P$_{\rm 1.4GHz} < 8.5 \times 10^{21}$W Hz$^{-1}$ (using a radio spectral index $\alpha = 0.5$). Note this upper limit applies to the combined galaxy+core, not just the central AGN. 

\section{Discussion}

The optical spectrum of 2QZJ215454.3$-$305654 is inconsistent with that of a  pure elliptical galaxy, while the lack of strong emission lines exclude it from being a nuclear starburst or a type I QSO; no combination of elliptical and starburst alone can reproduce the strong blue continuum of our source. Neither is this source likely to be a dust obscured type II AGN as the spectrum is blue rather than red. Galaxy fits using a ``diluted'' (1.5\%) starburst plus elliptical (4.5\%) plus power-law (94\%) resulted in a spectrum similar to that of 2QZJ215454.3$-$305654. The puzzle therefore is whether this object  is a virtually lineless AGN with the optical coninuum emission emanating from an accretion disk, or instead a radio-quiet/radio-weak object with optical synchrotron emission from a relativistic jet viewed at a small angle to the line of sight, i.e. a radio-quiet BL Lac object. Similar radio-quiet, X-ray weak, lineless QSOs have been found in the Sloan Digital Sky Survey (Fan et al.\,1999, Leighly et al.\,2004). Leighly et al.\,(2004) suggest that high-$z$ lineless QSOs might be early Universe counterparts of luminous narrow-line Seyfert 1 galaxies.
 
To establish the extent to which the lack of radio emission is unusual in an object that otherwise displays many characteristics of a BL Lac\footnote{defined in  The Astronomy and Astrophysics Encylopedia (ed. Maran) as `` . . . point-like sources of
optical radiation that show little or no line emission, and strong and
variable brightness and polarization."}
 we compare its radio-to-optical flux ratio ($\alpha_{\rm ro}$) with an unbiased sample of X-ray selected BL Lacs from the HRX\footnote{Hamburg-RASS X-ray selected sample of BL Lacs, Beckmann et al.\,2003} BL Lac survey (Beckmann et al.\ 2003).  In our calculations of $\alpha_{\rm ro}$ values we use flux densities in the observed frame to maintain consistency with values derived for the HRX BL Lacs.

Care needs to be taken in the calculation of $\alpha_{\rm ro}$ for a variable object, where
\begin{center}
{\large
$\alpha_{\rm ro} = - \frac{log \ f_r/f_o}{log \ \nu_r/\nu_o}$},
\end{center}
particularly when the radio and optical flux measurements are non-contemporaneous.  We therefore use the optical/radio observations separated by the smallest time interval; the $I$ band magnitude derived from the ESO CCD observations (August 2003) and the 1.4GHz 3$\sigma$ upper flux limit from the ATCA observations (November 2003). These observations also provide the most accurate limits on the flux of the object in both regimes.  However we note that 2QZJ215454.3$-$305654 has been observed to exhibit significant optical variability ($0.5\,$mag) on these time scales (see Fig.\,1).

We calculate $\alpha_{\rm ro}$ values based on both the total $I$ band magnitude of the object ($m_I = 18.83$) and the AGN (core) component in the case of the fainter elliptical galaxy fit ($m_{\rm core} =19.29$). This should bracket the likely range of $I$ band magnitudes for the AGN component. Indeed, we choose not to use the ratios derived from the spectroscopic fitting, in order to avoid slit loss corrections and to obtain a firm lower limit of the $I$-band magnitude of the core component.  

For $I_{\rm total}$ ($0.076\,$mJy) and $I_{\rm core}$ ($0.050\,$mJy) we obtain a 3$\sigma$ upper limit of $\alpha_{\rm ro}<0.047$ and  $\alpha_{\rm ro}<0.082$ respectively.  We note however that the image fitting is likely to have significant errors; from spectral fitting in \S2 we would expect a more dominant core component, thus computed optical core luminosities can  be treated as lower limits, resulting in an even more stringent upper limit on the calculated $\alpha_{\rm ro}$ value.

\begin{figure}
\psfig{file=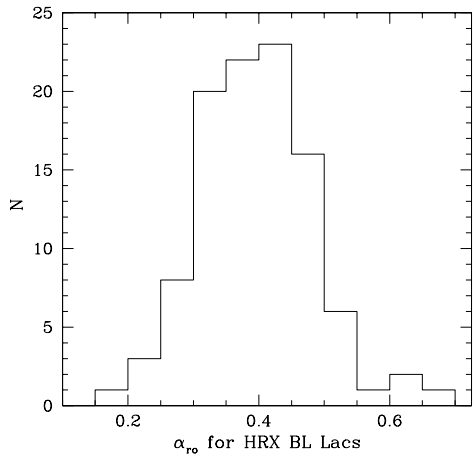,height=2.7in}
\caption{Histogram of $\alpha_{\rm ro} $ values for the 103 X-ray selected BL Lac objects from the HRX BL Lac survey (Beckmann et al.\,2003). One object has a value of 0.13 (subsequently revised to 0.34, see text for details), while the remaining 102 have $\alpha_{\rm ro} > 0.2$.}
\label{hrx}
\end{figure}

The distribution of $\alpha_{\rm ro}$ values for the HBX BL Lac sample (redshift range $0.030 < z < 0.89$) is shown in Fig.\,\ref{hrx}. The mean $\langle \alpha_{\rm ro} \rangle$ is 0.37, with a lowest computed value of $\alpha_{\rm ro}=0.13$. We note that the 1.4GHz and $B$-band measurements were not taken contemporaneously and that none of the $B$-band fluxes in the HRX BL Lac sample have been corrected for host galaxy contribution (Beckmann, private communication). 

Comparing first $\alpha_{\rm ro}$  values uncorrected for host galaxy contamination, we find that
2QZJ215454.3-305634 exhibits a radio-to-optical flux ratio that is more than 60 times lower than the mean value for
the HRX sample ($\alpha_{\rm ro (total)} < 0.05$ compared to $\langle \alpha_{\rm ro} \rangle=$0.37). The uncorrected  $\alpha_{\rm ro}$  values also imply that 2QZJ215454.3-305634 has a total
radio-to-optical flux ratio at least 3 times lower than
IES1255+244, the HRX BL Lac with the lowest $\alpha_{\rm ro}$  value ($f_{\rm 1.4GHz} = 14.7$mJy and $m_B = 15.4$). 
However in the case of IES\,1255+244 ($z=0.141$) there is  a clearly extended host galaxy on the $B$-band image; indeed 
Heidt et al.\,(1999) measured the core R-band (650nm) flux of this object to be $m_R = 17.9$ ($m_{b_J} \sim 18.37$) giving $\alpha_{\rm ro} = 0.34$ for this object.  Similar removal of host galaxy flux for another technically radio-quiet HRX object (Beckmann, private communication) raises the minimum observed  $\alpha_{\rm ro}$ value for the HBX to $> 0.23$.  With
the galaxy-corrected  $\alpha_{\rm ro}$ upper limit for 2QZJ215454.3$-$305654  of 0.08, this
translates to a 3$\sigma$ upper limit for the radio-to-optical flux ratio  that is at least 6--7 times lower than the lowest observed value in the HBX
sample.    The radio properties of this optically featureless, variable source are thus likely to be significantly different to those of BL Lac objects discovered to date.

2QZJ215454.3$-$305654 is clearly not a ``traditional'' BL Lac, but may instead belong to a hitherto unrecognised population of radio-quiet continuum objects.   Whether optical appearance alone is sufficient to classify an object as a BL Lac is a moot point.  The optical continuum emission from such an object could be radiation from the accretion disk, thus we are seeing a radio-quiet AGN; the absence of clear emission lines in the spectrum may result from a lack of gas clouds illuminated by the source, or the ionizing radiation from the source may have been absorbed or scattered by dust.  This object might be similar to objects in an X-ray selected sample of Seyfert 2 type galaxies studied by Nicastro et al.\,(2003)  in which instability patterns in the accretion disk (rather than dust obscuration) result in the lack of a  broad line region, although not of emission lines from the narrow line region. Alternatively Leighly et al.\,(2004) suggest that lineless QSOs result from a very high accretion rate and a strong UV-peaked continuum. If, however, the optical emission of 2QZJ215454.3$-$305654 is beamed synchrotron radiation then the absence of photons at radio energies needs to be explained. Velocity structures in the jet (proposed by Chiaberge et al.\,2000) resulting in higher kinematic Doppler factors for optical photons in the spine of the jet, might be one such explanation. Deep X-ray observations of 2QZJ215454.3$-$305654 would provide further clues as to the identity of this object.

2QZJ215454.3$-$305654 is only one of $\sim$30 objects, first identified in the 2QZ/6QZ surveys, that display similar featureless 'nonthermal' optical spectra with no associated radio emission.  Ongoing detailed observations of these objects may demonstrate that 2QZJ215454.3$-$305654 is not alone, and that a population of hitherto unrecognised radio-weak ``BL Lacs'' makes up yet another exhibit in the AGN zoo. 

\section*{Acknowledgments}
We thank the anonymous refereee for his/her insightful comments which helped
greatly improve the contents of the paper.\\
The 2QZ/6QZ is based on observations made with the Anglo-Australian
Telescope and the UK Schmidt Telescope; we would like to thank our
colleagues on the 2dF and 6dF survey teams and all the staff at
the AAT who have helped make this survey possible. We also thank
the staff at the Mt Stromlo Observatory (observations of 2BL objects in October 2002 were among the last observations made with the 74'' telescope before it was desroyed in the fires of January 2003) and at the Australia Telescope Compact Array, which is funded by
 The Commonwealth of Australia for operation as a national Facility 
managed by CSIRO.\\
  DL thanks the School of Physics at the
University of Sydney for a Postgraduate (Mature Age) Scholarship.\\
JH acknowledges support by the Deutsche
  Forschungsgemeinschaft (SFB 439).\\

\label{lastpage}

\end{document}